# Where are your eyes going to look during reading? A critical evaluation of saccade targeting hypothesis

Yanping Liu[1] & Huan Wei[2]

[1]Key Laboratory of Behavioral Science

Institute of Psychology, Chinese Academy of Sciences

[2]Department of Psychology, Sun Yat-Sen University, China

Author Note

This research was supported by the grants from the China Postdoctoral Science Foundation (2013M541073 & 2014T70132). Correspondence should be addressed to Yanping Liu, 16 Lincui Road, Key Laboratory of Behavioral Science, Institute of Psychology, Chinese Academy of Sciences, Beijing, China. Email: liuyp@psych.ac.cn.

**Abstract**

The word-based account of saccades drawn by a central gravity of the PVL is supported by two pillars of evidences. The first is the finding of the initial fixation location on a word resembled a normal distribution (Rayner, 1979). The other is the finding of a moderate slope coefficient between the launch site and the landing site ($b$=0.49, see McConkie, Kerr, Reddix, & Zola, 1988). Four simulations on different saccade targeting strategies and one eye-movement experiment of Chinese reading have been conducted to evaluate the two findings. We demonstrated that the current understanding of the word-based account is not conclusive by showing an alternative strategy of the word-based account and identifying the problem with the calculation of the slope coefficient. Although almost all the computational models of eye-movement control during reading have built on the two findings, future efforts should be directed to understand the precise contribution of different saccade targeting strategies, and to know how their weighting might vary across desperate writing systems.

Visual system is constrained by inhomogeneous visual acuity across the retina with the fovea as the most acute part. During reading, it is well believed that, to compensate this limitation, the fovea is constantly directed to an optimal position. In line with it, it has been observed that the initial fixations on a word appear a normal distribution (i.e., the initial fixation location distribution, IFLD), the peak of which lies somewhere around the center of the word (e.g., Deutsch & Rayner, 1999; Dodge, 1906; Dunn-Rankin, 1978; Rayner, 1979; Rayner, Sereno, & Raney, 1996; Reingold, Reichle, Glaholt, & Sheridan, 2012; Vitu, O'Regan, & Mittau, 1990). According to the word-based account, this point has been referred as the *preferred viewing location* (PVL), which can prioritize ongoing lexical processing (Rayner, 1979).

To further explain the generation of the IFLD, especially the PVL, the range effect has been found to indicate a central gravity of the PVL to draw eyes (e.g., McConkie et al., 1988). The range effect describes a pattern of variation of the saccade length as the launch sites (i.e., prior fixations) approaching the center of the next words. In another word, saccades tend to undershoot or overshoot the next word center when eyes are far away from or close to it, respectively.

Almost all the computational models of eye-movement control during reading have incorporated the PVL and the range effect as the mechanism of saccade targeting (e.g., *Bayesian reader*: Engbert, & Krügel, 2010; *SWIFT*: Engbert, Nuthmann, Richter, & Kliegl, 2005; *E-Z Reader*: Reichle, Pollatsek, Fisher, & Rayner, 1998; Reichle, Rayner, & Pollatsek, 1999; *Glenmore*: Reilly & Radach, 2006). However, the purpose of the present note is trying to evaluate the word-based account by offering an alternative strategy that produces the similar IFLD and identifying the problem of the range effect with its method of calculation.

**An alternative strategy of saccade targeting in producing the similar IFLD**

The word-based account in general states that the adjustment of saccade targeting to an optimal position is influenced by the information of word length. Logically speaking, there are two possible loci—either on the landing word or on the launch word— being the candidates of optimal positions. Therefore, to explain the generation of the PVL, saccades could converge to that point directly or alternatively, be tuned to an optimal position on the launch word before a move of constant length to that point. The first strategy, which is widely accepted, can be seen as a direct instantiation of the word-based account.

The alternative strategy, which is what we propose, assumes that the adjustment takes place at the launch site instead of the landing site. The eyes re-fixate to an optimal site on the launch word by an intra-word saccade before an inter-word saccade of constant length to the next word. Consequently, saccades under this strategy may coincidentally locate around the center of the next word, generating the similar IFLD. Thus, this strategy can be seen as an indirect instantiation of the word-based account.

To test the alternative strategy, four models were built based on different assumptions of saccade targeting strategies in a writing system assembled as the alphabetic language (i.e., spaced). During simulations, each model was run to "read" 100,000 artificial sentences, each of which consisted of 10 different words. The lengths of the first and last words in each sentence maintained 1, and the lengths of the other words were random permutations between 2 and 9. For simplicity, the same Greek symbols in the following formulae represent the parameters for different models. Their corresponding values are listed in Table 1.

Model I assumes that eyes move in a constant saccade length. Parameter φ is a constant representing the minimal saccade length. The inter-word saccade length, or INTER-SL is given by:

INTER-SL = φ, with a Gaussian noise $N(0, \sigma)$ for saccade error        (1)

Model II and III assume that readers adjust their saccade length based on the distance between the launch site and the center of the next word. For Model II, the saccade length becomes longer when the launch site approaching the center of the next word. In this case, the landing site tries to move away from the center of the next word, so the slope coefficient between the launch site and the landing site is large ($b > 1$). For Model III, the saccade length becomes shorter when the launch site approaching the center of the next word, so the slope coefficient between the launch site and the landing site is moderate ($0 < b < 1$). Model III is exactly the same mechanism as McConkie et al.'s to adjust the landing site. D represents the distance between the launch site and the center of next word. β is a parameter that modulates this distance. The inter-word saccade length, or INTER-SL, for Model II and III is given by:

INTER-SL = φ + β × D, with a Gaussian noise $N(0, \sigma)$ for saccade error   (2)

Model IV assumes that readers adjust the launch site by making an intra-word saccade and then making an inter-word saccade of constant length. This model is the alternative strategy proposed by this note. E denotes the eccentricity acuity, or the center of the fixated word subtracting the launch site. π and λ are parameters that modulate the influence of visual acuity on the probability of intra-word saccade.

The intra-word saccade probability, or INTRA-SP is given by:

INTRA-SP = $(|\pi E|)^\lambda$                                               (3)

The intra-word saccade length, or INTRA-SL is given by:

INTRA-SL = E, with a Gaussian noise $N(0, \varsigma)$ for saccade error     (4)

The inter-word saccade length, or INTER-SL is given by:

INTER-SL = φ, with a Gaussian noise $N(0, \sigma)$ for saccade error     (5)

Figure 1 shows the IFLD for each model. In Model I, the IFLD was flat at first and suddenly dropped after some positions (Figure. 1a). In Model II, the IFLD decreased slightly at first and suddenly dropped after some positions (Figure. 1b). In Model III and IV, both IFLD resembled the normal distribution (Figure. 1c-d), meaning that the saccade pattern produced by Model IV, the alternative strategy, was also consistent with the qualitative patterns of eye movement observed during alphabetic reading.

**The existing problem of range effect**

The range effect is essential to support the direct instantiation of word-based account (McConkie et al., 1988). As stated above, it reflects the variation of the saccade length as the function of the distance between the launch site and the center of the next word. It is indicated by a moderate slope coefficient ($0 < b < 1$) between the landing site and the launch site. However, the traditional way to calculate this coefficient is problematic because of data selection. Due to the presupposed word-based account, the calculation used incoming saccades, the saccades that shot exactly on the target word after they launched from the word prior to that target word. Conversely, we propose to use outgoing saccades which are any forward saccades that launch from words immediately preceding the target words, including not only shot but also undershot and overshot at the target word.

Figure 2 shows the coefficients between the launch site and the landing site of the above four models for incoming and outgoing saccades, respectively. For outgoing saccades, only in Model III the slope coefficient was within a moderate range ($0.48 \leq$ $b \leq 0.50$; see Table 2). For incoming saccades, however, the slope coefficients maintain moderate for all four models ($0.15 \leq b \leq 0.79$; see Table 2). Therefore, the range effect based on the slope coefficient calculated from incoming saccades is insufficient to discriminate different saccade targeting strategies.

Figure 3 also shows the relationship between the launch site and the landing site for incoming and outgoing saccades but from an eye-movement experiment of Chinese reading (Liu, Li, & Pollatsek, submitted for review).While the slope coefficient for outgoing saccades was larger than 1, for incoming saccades, it was still within a moderate range ($b = 0.42$). In a language system that has no specific PVL due to the absence of blank spaces between words (see Liu, Li, & Rayner, 2011; cf., Yan, Kliegl, Richter, Nuthmann, & Shu, 2010), the persistent range effect leads to an obvious contradiction.

## Discussion and Conclusion

The aim of this note is to evaluate the word-based account and their corresponding evidences. We have demonstrated that the central tendency of IFLD can be produced by direct and indirect instantiation of the word-based account equivalently. It should be noted that both strategies are not necessary to mutually exclusive, the selection of strategy may depend on the certainty for perceiving the center of next word. As a result, it is better to weight both strategies to account for the complicated pattern of saccade targeting behaviors during reading. Moreover, we have shown that the slope coefficients of incoming saccades maintain moderate, irrespective of the underlying saccade targeting strategies. Thereby, the range effect calculated from the traditional way is unable to confirm or deny any specific mechanism of saccade targeting.

The intra-word refixation saccades exist broadly during reading (e.g., the U-shaped function of refixation probability; McConkie, Kerr, Reddix, Zola, & Jacobs, 1989; Rayner et al., 1996, see Figure 3). We should reconsider the role they might play. In the alternative strategy, when the launch site is far away from the center of the next word, it is likely to make a rightward refixation on the prior word. Conversely, when the launch site is close to the center of the next word, it is likely to make a leftward refixation on the prior word. It is possible during real reading that the summation of the intra-word saccades and the subsequent approximately constant inter-word saccades acts like the range effect to produce the central tendency of the IFLD on the next word. If then, the intra-word saccades may not only contribute to process the current word but also assist eyes to target on the PVL of the next word for further processing.

During reading, due to systematic and random errors in the oculomotor system, there are rather broad and overlapping tails between the distributions of within-word landing positions (e.g., Engbert & Nuthmann, 2008; McConkie et al., 1988). The tails cannot be ignored and excluded arbitrarily as undershot or overshot at the target word. Otherwise, the central tendency of the target word will always exhibit. We believe that the use of outgoing saccades to calculate the range effect is more promising to discriminate different saccade targeting strategies. For instance, the slope coefficient of outgoing saccades in model II was larger than 1. The strategy underling this model is consistent with the converged evidences from studies of Chinese reading in which saccade length increases with more parafoveal processing (Liu, Reichle, & Li, 2014; Liu et al., submitted for review). Meanwhile, the slope coefficient of outgoing saccades in model IV was moderate, given the underlying assumption that eyes were guided to the center of the next word directly.

In summary, our evaluation raises question about the word-based account and the underlying models. Further empirical and modeling works should attend to figure out which strategy or even a combination of them is a better account of eye movement control during reading, and to know their abilities of adaption to disparate writing systems.

isolated words and continuous text. *Perception and Psychophysics, 47*, 583-600.

*Table 1*. Parameters defining inter-word saccades and intra-word saccades in the four models, and their corresponding values

| Model | Inter-word saccades | | | Intra-word saccades | | |
|---|---|---|---|---|---|---|
| | $\varphi$ | $\beta$ | $\sigma$ | $\pi$ | $\lambda$ | $\varsigma$ |
| I | 7 | - | 1.5 | - | - | - |
| II | 8 | -0.3 | 1.5 | - | - | - |
| III | 3.5 | 0.5 | 1.5 | - | - | - |
| IV | 7 | - | 1.5 | 0.2 | 1.05 | 0.5 |

*Table 2*. The slope coefficients *b* between the launch site and the landing site when fitting the incoming and outgoing saccades from the four models (the target word length is from 2 to 9, respectively).

| Model | Incoming Saccades | | | | | | | | Outgoing Saccades | | | | | | | |
|---|---|---|---|---|---|---|---|---|---|---|---|---|---|---|---|---|
| | 2 | 3 | 4 | 5 | 6 | 7 | 8 | 9 | 2 | 3 | 4 | 5 | 6 | 7 | 8 | 9 |
| I | 0.25 | 0.38 | 0.49 | 0.60 | 0.68 | 0.74 | 0.77 | 0.79 | 0.98 | 0.98 | 0.98 | 0.99 | 0.99 | 1.00 | 1.00 | 1.00 |
| II | 0.27 | 0.40 | 0.53 | 0.63 | 0.70 | 0.74 | 0.75 | 0.73 | 1.29 | 1.29 | 1.29 | 1.29 | 1.29 | 1.30 | 1.30 | 1.29 |
| III | 0.15 | 0.24 | 0.31 | 0.39 | 0.45 | 0.49 | 0.53 | 0.57 | 0.48 | 0.49 | 0.49 | 0.49 | 0.50 | 0.50 | 0.50 | 0.50 |
| IV | 0.25 | 0.38 | 0.50 | 0.60 | 0.69 | 0.74 | 0.76 | 0.77 | 1.33 | 1.34 | 1.32 | 1.32 | 1.31 | 1.30 | 1.31 | 1.36 |

**Figure Captions**

*Figure 1*. Simulation of the frequency distribution of initial fixation location on the target word for the four models. In Model I, the saccade length was constant; in Model II, the saccade length became longer when approaching the center of next word; in Model III, the saccade length became shorter when approaching the center of next word; in Model IV, eyes made an intra-word saccade to the center of the fixated word with a probability, then made a constant inter-word saccade to the right. All saccades in these models have Gaussian saccade errors. The abscissas shows the character position on the target word, with 0 indicating the blank space immediately to the left of the target word.

*Figure 2*. The saccade landing site as a function of the launch site. The graph shows the simulated data and regression lines fitted to incoming saccades and outgoing saccades. The launch site and the landing site were aligned to the center of the target word (length = 5). The ordinate value represents the position of saccade targeting relative to the target word. 0 is on the word center, -3 to 3 is within the word (light gray diamonds), less than -3 is undershot of the word (lower gray triangular), and more than 3 is overshot of the word (upper gray triangular). The abscissa value represents the distance from the launch site on the prior word to the center of the target word. The four models were defined identical with those in Figure 1.

*Figure 3*. The saccades landing site as a function of the launch site during Chinese reading. The graph shows the empirical data and their regression lines fitted to incoming saccades (i.e., -1 < landing site < 1) and outgoing saccades (all data points), respectively. The launch site and the landing site for incoming saccades are aligned to the center of the target word (2 characters), but for outgoing saccades, they are aligned to the center of the word immediately following the target word (also 2

characters). The ordinate value represents the position of saccade targeting relative to the word immediately following the target word. 0 is on the word center, -1 to 1 is within the word (light gray diamonds), less than -1 is undershot of the word (lower gray triangular), and more than 1 is overshot of the word (upper gray triangular). We rendered the incoming regression lines on the outgoing saccades for illustrative purpose only.

*Figure 1.*

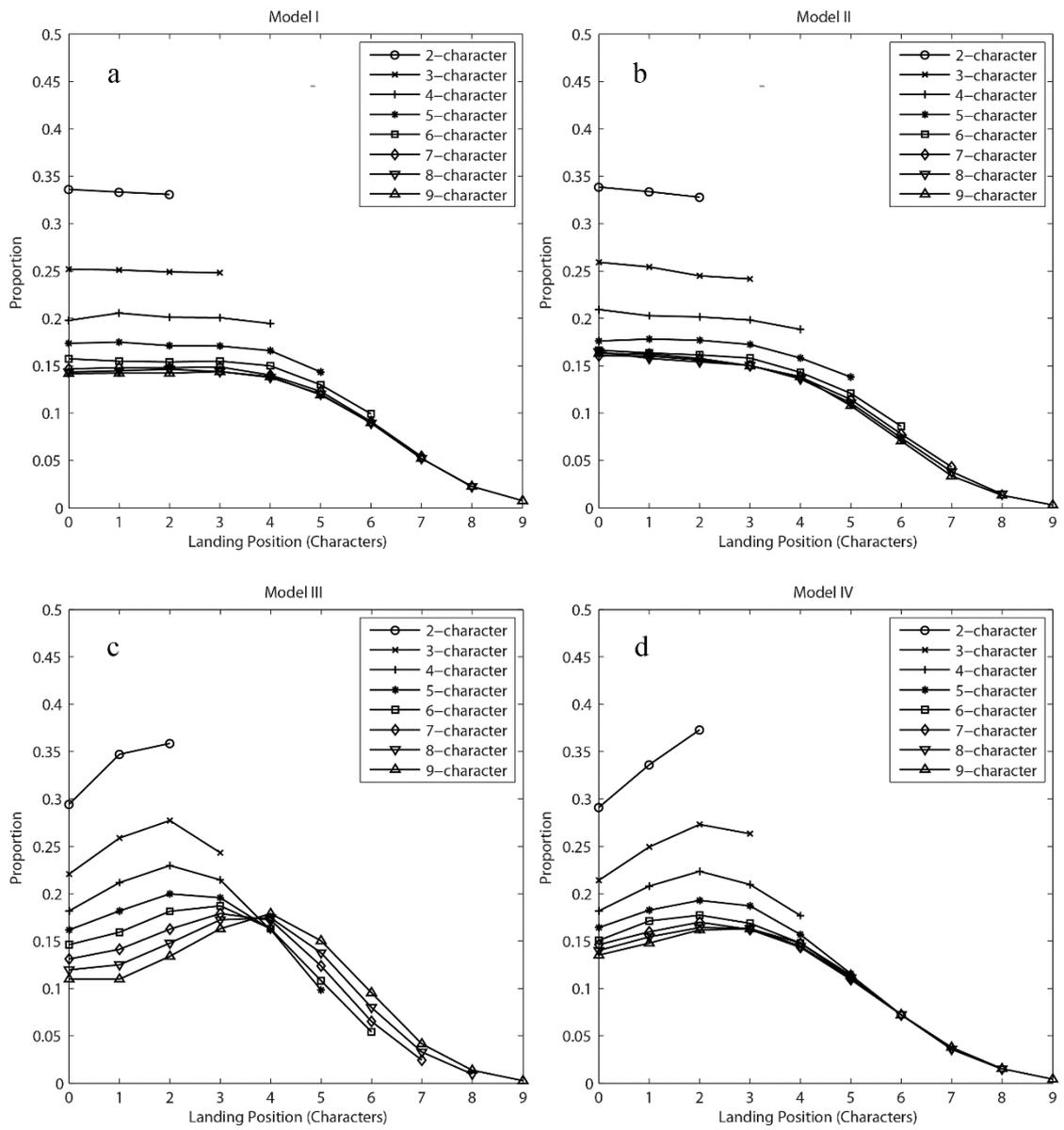

*Figure 2.*

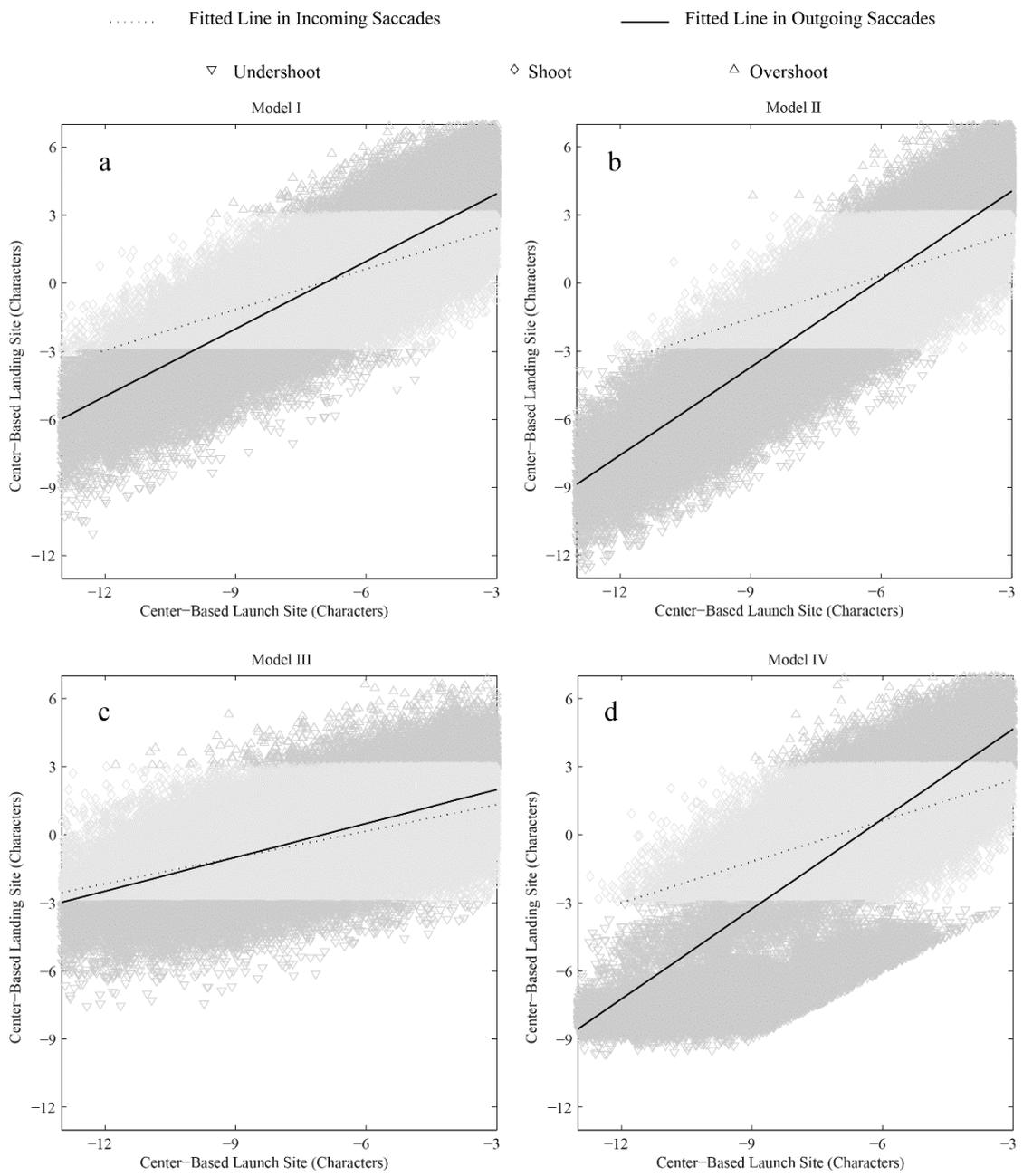

*Figure 3.*

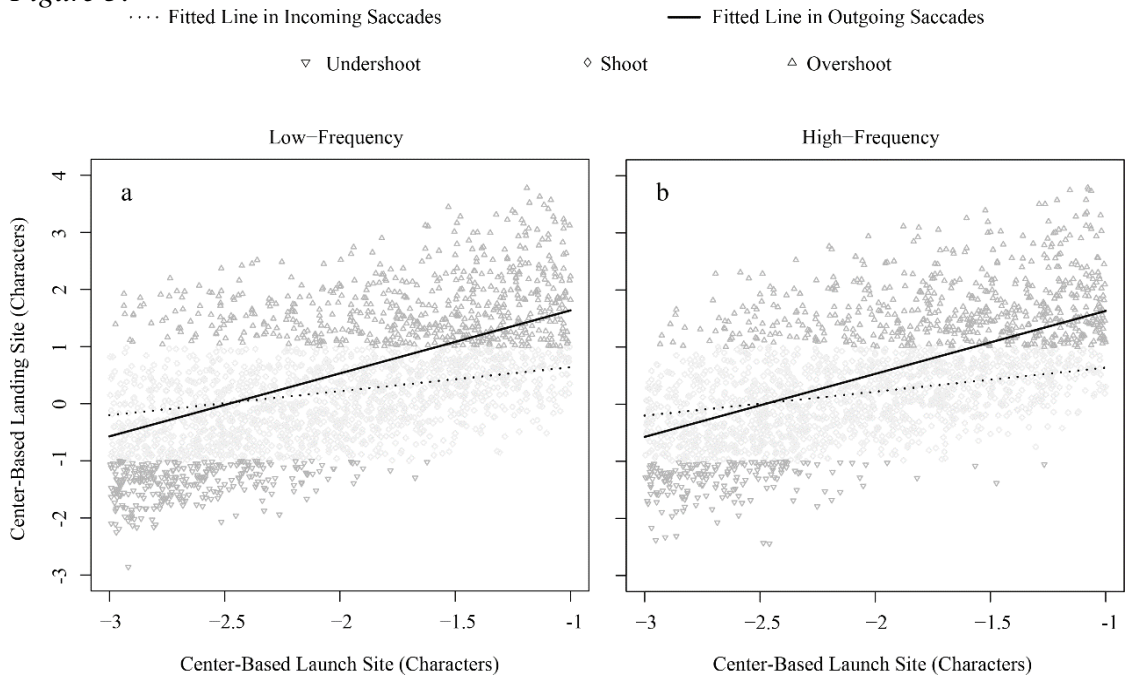